# THE GEOMETRY OF STATIONARY SPACETIMES (SUCH AS THE KERR SOLUTION) THAT ARE ASYMPTOTICALLY STATIC AT INFINITY

John E Heighway

The present work is an extension of the paper entitled "Towards a deeper Understanding of General Relativity," paper number gr-qc/0001036 [1]. In the referenced paper, it is shown that ordinary GR admits an extended interpretation as a variable rest mass theory. As argued in that paper, the dependence of rest mass upon gravitational potential affects not only the rate of clocks (the so-called time dilation effect) but also results in the elongation of measuring rods. In view of these circumstances, an alternate system of measurment that is unaffected by gravity is introduced, resulting in a new metric which is conformally related to the usual proper metric of GR. Evidence is presented to show that the new "telemetric" system of measurment results in a geometry has an excellent claim to be the true geometry of the Schwarzschild solution. The purpose of the present paper is to extend the Variable Rest Mass (VRM) Interpretation and the telemetric system of measurment to the case of stationary but non-static spacetimes, especially the Kerr solution.

## Introduction

The approach taken here is to separate dynamical (frame-dragging) effects from those that may be thought of as purely gravitational effects.
For each point in the field, a velocity is calculated which maximizes the rate of a clock moving with that velocity. The residual slowing of such clocks is assumed to be purely gravitational and, in accord with the VRM Interpretation, the concomitant length dilation effect fixes the relation between the usual geometry defined by proper measurements, on the one hand, and the telemetric geometry, on the other.

In order to test the theory, the method is applied to the Kerr solution, and the results are used to calculate the location of the famous unstable photon orbits. The orbits are calculated by assuming that the they will lie on circles whose (telemetric) curvature matches that induced in the ray paths by the shear in the frame-dragging velocity. The results are exactly those calculated directly by the usual methods.

## Geometry and Frame Dragging

In the paper [1] referenced in the introduction, it is emphasized that the telemetric system of measurement has as its basis the fact that in any stationary field the frequency of light is constant along its path. That paper considered only the Schwarzschild field, which is not just stationary, but static. The Telemetric system can be extended to fields that are stationary but non-static, provided that they are asymptotically static at infinity. This condition is necessary in order to insure that a single time standard can be propagated to all observers.

In non-static stationary fields, the metric tensor contains mixed space-time components which cannot be eliminated except at the cost of introducing explicit time dependence. In such situations, one must disentangle geometric effects from the dynamical effects associated with the so-called 'frame-dragging' phenomenon. The Kerr solution provides an excellent example of this. Here space itself seems to be moving in circles in a vortex-like manner about the rotating object, as evidenced by the fact that light rays reflected by a series of mirrors arranged on a circle centered on the axis of rotation will require less time when directed with the rotation as compared with rays directed oppositely. Furthermore, in the limit of infinitely many mirrors, i.e., for a circular path, the results are compatible with an interpretation, which assigns an azimuthal velocity, v, to that which must remain nameless for fear of using the ae-word, such that the circuit times are inversely proportional to c+v and c–v respectively. Furthermore, and this is a crucial point, the rate of a clock is maximized when it 'goes with the flow,' i.e., when it moves with the same azimuthal velocity, v. Finally, the slowing of stationary proper clocks by the factor $\sqrt{[1 - (v/c)^2]}$ in relation to the maximum rate can be understood in terms of the special relativity time dilation effect.

The telemetric interpretation understands the time dilation of the fastest, 'go-with -the-flow' clocks as resulting from the rest-mass reduction effect, and the concomitant length dilation effect fixes the relationship between telemetric lengths and proper lengths in the v-moving frame. These considerations allow one to disentangle the dynamical and geometric effects.

## Finding the Fastest Clocks

For the general case, one must determine the velocity, v, at each point in space-time such that the rate of a clock moving with that velocity is a maximum. The rate of a clock relative to world time is $ds/dx^0$ where

$$(ds/dx^0)^2 = g_{00} + 2g_{0\alpha}(dx^\alpha/dx^0) + g_{\alpha\beta}(dx^\alpha/dx^0)(dx^\beta/dx^0)$$

Here regarding sub- or super-scripts we use Greek for space indices (1,2,3) and Latin for space-time (0,1,2,3). Also the sign of our $g_{ik}$ are reversed as compared with Landau and Lifshitz.

Setting the derivative of $(ds/dx^0)^2$ with respect to $(dx^\alpha/dx^0)$ equal to zero yields the system of equations for the motion $dx^\alpha/dx^0$ which will maximize the rate of a clock. The system of equations which results simplifies to [A1]

$$g_{\alpha\beta}\, dx^\beta/dx^0 = -g_{0\alpha}$$

The inverse of the pure space submatrix $g_{\alpha\beta}$ is [2]

$$g^{\alpha\beta} - g^{0\alpha} g^{0\beta}/g^{00}$$

The solution to the system is thus [A2]

$$dx^\alpha/dx^0 = -(g^{\alpha\beta} - g^{0\alpha} g^{0\beta}/g^{00})\, g_{0\beta} = g^{0\alpha}/g^{00}$$

Inserting this result into the expression for $(ds/dx^0)^2$ one finds [A3]

$$(ds/dx^0)^2 = 1/g^{00}$$

For the Schwarzschild metric, $1/g^{00} = g_{00}$. Our result is thus consonant with the results of reference [1].

## The Frame Dragging Velocity

The contravariant components of the flow velocity are defined by [3]

$$\beta^\alpha := v^\alpha/c := (dx^\alpha/dx^0)/\{\sqrt{g_{00}}[1+(g_{0\alpha}/g_{00})(dx^\alpha/dx^0)]\}$$

which reduces to [A4]

$$\beta^\alpha = \sqrt{(g_{00})}\, g^{0\alpha}$$

The covariant components are obtained using the pure space metric tensor [4]

$$\lambda_{\mu\nu} = -g_{\mu\nu} + g_{0\mu}g_{0\nu}/g_{00}$$

Thus [A5]

$$\beta_\alpha = \lambda_{\alpha\mu}\beta^\mu = g_{0\alpha}/\sqrt{(g_{00})}$$

The physical velocity has magnitude $\beta$ where

$$\beta^2 = \beta_\alpha \beta^\alpha = g_{0\alpha} g^{0\alpha} = 1 - g_{00} g^{00}$$

and finalIy, the gamma factor for the fIow is

$$\gamma = 1/\sqrt{(1-\beta^2)} = 1/\sqrt{(g_{00}g^{00})}$$

## The Telemetric Line Element

Next we shall introduce what may be called 'go-with-the-flow' coordinates $\underline{x}^\alpha$ satisfying

$$d\underline{x}^\alpha = dx^\alpha - g^{0\alpha}/g^{00} \, dx^0$$

Inserting these reduces the invariant line element to [A6]

$$ds^2 = 1/g^{00} (dx^0)^2 + g_{\alpha\beta} d\underline{x}^\alpha d\underline{x}^\beta$$

From this we can deduce the telemetric line element, which uses world time, and bases its distance measurements on echo ranging using world time. Thus $g^*_{00} = 1$, and $\lambda^*_{\alpha\beta} = -g^{00}g_{\alpha\beta}$, and we can write

$$ds^{*2} = (dx^0)^2 + g^{00} g_{\alpha\beta} \, d\underline{x}^\alpha d\underline{x}^\beta = g^{00} \, ds^2$$

Thus the conformal relationship between telemetric and proper metrics holds in the more general case, provided that one properly takes account of the dynamical effects of 'frame dragging.'

## The Kerr Solution

For the Kerr solution these results may be listed using the usual coordinates and abbreviations

$$r_s := 2GM/c^2, \quad a := L/Mc, \quad L = \text{angular momentum}, \quad M = \text{mass}$$

$$\rho^2 := r^2 + a^2 \cos^2\theta, \quad \text{and} \quad \Delta := r^2 + a^2 - r_s r$$

along with $u^4 := \rho^2(r^2 + a^2) + r_s r a^2 \sin^2\theta$, as follows:

$$d\phi/dx^0 = r_s r a/u^4, \qquad \beta = r_s r a \sin\theta/(\rho^2 \sqrt{\Delta})$$

$$\gamma = (\rho^2 \sqrt{\Delta})/[\sqrt{(\rho^2 - r_s r)}u^2]$$

$$(d\tau/d\tau*)_{max} = 1/\sqrt{g^{00}} = (\rho\sqrt{\Delta})/u^2$$

This last form is equal to the mass-reduction factor, i.e.,

$$m*_e/m_e = 1/\sqrt{g^{00}} = (\rho\sqrt{\Delta})/u^2$$

and thus the event horizon occurs where $\Delta = 0$. Thus there are two such surfaces, the 'spheres' ( r is only a parameter) defined by

$$r_h = r_s/2 \{1 \pm \sqrt{[1 - (2a/r_s)^2]}\}$$

The region between these horizons seems unphysical, since in that region, the coordinate r becomes timelike while t becomes spacelike. The region for which r falls in the range between zero and the smaller horizon value may represent another universe which seems not to have been investigated.

For values of the parameter, a , greater than $r_s/2$, both horizons vanish, and this exotic universe (Shangri-La ?) merges with the innerspace of our universe.

The metric forms of interest are

stationary

$$ds^2 = (\rho^2 - r_s r)\rho^{-2} (dx^0)^2 + 2r_s r a \sin^2\theta \rho^{-2} dx^0 d\phi$$
$$- u^4 \rho^{-2} \sin^2\theta \, d\phi^2 - \rho^2/\Delta \, dr^2 - \rho^2 d\theta^2$$

go-with-the-flow [A7]

$$ds^2 = \rho^2 \Delta/u^4 (dx^0)^2 - u^4 \rho^{-2} \sin^2\theta \, d\underline{\phi}^2 - \rho^2/\Delta \, dr^2 - \rho^2 \, d\theta^2$$

telemetric

$$ds*^2 = (dx^0)^2 - u^8/(\rho^4 \Delta) \sin^2\theta \, d\underline{\phi}^2 - u^4/\Delta^2 \, dr^2 - u^4/\Delta \, d\theta^2.$$

In order to demonstrate the heuristic value of the telemetric interpretation, we shall investigate the two unstable photon orbits which occur in the Kerr solution. We need first to define the functions r* and R*. Let

$$r^* := u^4 / (\rho^2 \sqrt{\Delta}) \quad \text{and} \quad dR^*/dr := u^2/\Delta$$

so that the element of telemetric length in the $\phi$ and r directions are $r^* \sin\theta \, d\phi$, and $dR^*$, respectively.

## The Photon Orbits

Regarding the photon orbits, one would expect that in the outer region, beyond the stenosurface, wavefronts propagating in the counter-flow direction would be turned inward because of the shear in the flow. The curvature, $\kappa$, of the ray is easily calculated.

Referring to the embedding diagram shown in figure 1, one has

$$\rho^*/r^* = dR^*/dr^*$$

whence the telemetric curvature of a circle in the equatorial section is

$$\kappa_o = 1/\rho^* = (1/r^*)(dr^*/dR^*) = d/dR^*[\ln(r^*)]$$

From figure 2, depicting the (flattened) cone tangent to the embedding surface of figure 1, one has two equal expressions for the angle

$$[c - v(R^* - \Delta R^*)]\Delta t / (\kappa^{-1} - \Delta R^*) = [c - v(R^* + \Delta R^*)]\Delta t / (\kappa^{-1} + \Delta R^*)$$

Expanding v in a Taylors series, one has for $\Delta R^* \to 0$,

$$\kappa = -(dv/dR^*)/(c-v) = d/dR^*[\ln(c-v)]$$

If the telemetric picture correctly deals with dynamics and geometry, then the photon orbit should occur when these curvatures match, implying

$$d/dR^*[r^*/(c-v)] = 0$$

Now $2\pi r^*/(c-v)$ is just the telemetric transit time for a light signal traversing the ring of mirrors, and one would expect that this circle would correspond to an orbit when the transit time is a minimum. But expectations aside, it turns out that the above equation is in fact satisfied by the actual photon orbit as determined in the usual way, as will now be shown.

For the region inside the stenosurface, a wavefront propagating in the same direction as the rotation will be turned so as to circle that surface. In this case the curvatures are

$$\kappa = d/dR^*[\ln(c+v)] \quad \text{and} \quad \kappa_o = d/dR^*[\ln(r^*)]$$

so that the orbit should satisfy $d/dR^*[r^*/(c+v)] = 0$

The two cases will be treated simultaneously. With the benefit of hindsight, one may simplify the work by writing

$$r^*/(1 \mp \beta) \;=\; r^*(1 \pm \beta)/(1-\beta^2) \;=\; (\rho^2\sqrt{\Delta} \pm r_s\, r\, a\, \sin\theta)/(\rho^2 - r_s r)$$

The orbits lie in the equatorial plane, so that $\rho = r$. Also, $R^*$ is a monotone function of r. We may thus write the equations as

$$d/dr\,[(r\sqrt{\Delta} \pm r_s a)/(r - r_s)] = 0$$

This results in the following equation, valid for both orbits [A8]

$$4\,r^3 - 12\,r_s\,r^2 + 9\,r_s^2\,r - 8\,r_s\,a^2 = 0.$$

Setting $A = 2a/r_s$, the solutions may be written, first for the outer, retrograde orbit,

$$r_{ph-}/r_s \;=\; 1 + \cos(2/3\ \mathrm{acos}(A))\,, \qquad\qquad \text{for } A < 1\,,$$
$$\qquad\quad = 1 + \cosh(2/3\ \mathrm{acosh}(A))\,, \qquad\qquad \text{for } A > 1\,,$$

and for the "inner" (smaller values of r), prograde orbit,

$$r_{ph+}/r_s = 1 + \cos(2/3\ \mathrm{acos}(-A))$$

These are the well-known (unstable) photon orbits [5]. The third root,

$r_{ph+?}/r_s = 1 - \cos(1/3 \, \text{acos}(1 - 2A^2))$

is a prograde orbit in the 'Shangri-La' universe mentioned earlier.

Figure 3 shows the embedding diagram for the equatorial section of the Kerr geometry for three values of the rotation parameter $A = 2a/r_s$ (labeled $\alpha$ in fig 3). The heavy dots indicate the location of the photon orbits.

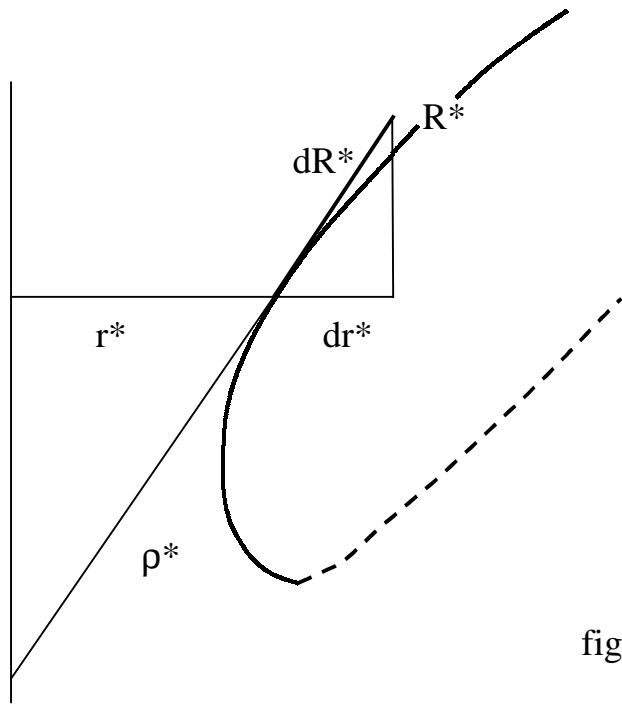

figure 1

$\rho^*/r^* = dR^*/dr^* \qquad \kappa_0 = 1/\rho^* = (1/r^*)(dr^*/dR^*) = d/dR^*[\ln(r^*)]$

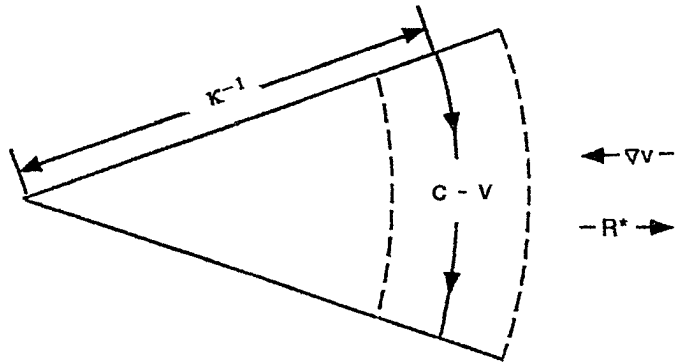

figure 2

$$[c-v(R*-\Delta R*)] \Delta t / (\kappa^{-1}-\Delta R*) = [c-v(R*+\Delta R*)] \Delta t / (\kappa^{-1}+\Delta R*)$$

for $\Delta R* \to 0$,    $\kappa = -(dv/dR*)/(c-v) = d/dR*[\ln(c-v)]$

$$[\kappa_0 = \kappa] \implies d/dR*[r*/(c-v)] = 0$$

---

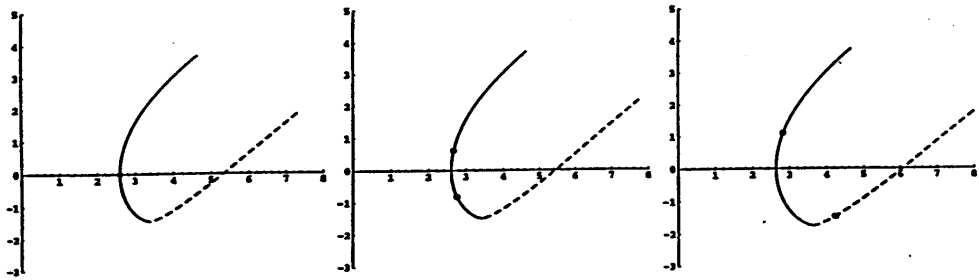

fig. 3. Telemetric imbedding diagrams for the equatorial section of the Kerr field for three values of the rotation parameter ($\alpha$ = 0.0, 0.2, 0.4), showing the location of the unstable photon orbits. Scale in units of $r_s$.

# Appendix: Details of Various Calculations

[A1]   Page 4.2    $g_{\alpha\beta} \, dx^\beta/dx^0 = -g_{0\alpha}$

In the expression for $(ds/dx^0)^2$ the final term will contain $g_{\alpha\alpha}(dx^\alpha/dx^0)^2$ along with two identical terms linear in $dx^\alpha/dx^0$ which together produce $2 \sum_{\beta\neq\alpha} g_{\alpha\beta}(dx^\beta/dx^0)(dx^\alpha/dx^0)$. Differentiating with respect to $dx^\alpha/dx^0$ thus yields

$2 g_{0\alpha} + 2 g_{\alpha\alpha} \, dx^\alpha/dx^0 + 2 \sum_{\beta\neq\alpha} g_{\alpha\beta}(dx^\beta/dx^0) \;=\; 2 g_{0\alpha} + 2 g_{\alpha\beta} \, dx^\beta/dx^0$

---

[A2]   Page 4.3    $dx^\alpha/dx^0 = g^{0\alpha}/g^{00}$

$dx^\alpha/dx^0 = -(g^{\alpha\beta} - g^{0\alpha} g^{0\beta}/g^{00}) \, g_{0\beta}$

$\qquad\qquad = -[(g^{\alpha k} g_{0k} - g^{\alpha 0} g_{oo}) - g^{0\alpha}/g^{00}(g^{0k} g_{0k} - g^{00} g_{00})]$

$\qquad\qquad = -[(\delta^\alpha_{\;0} - g^{0\alpha} g_{00}) - g^{0\alpha}/g^{00}(\delta^0_{\;0} - g^{00} g_{00})]$

$\qquad\qquad = -[(-g^{0\alpha} g_{00}) - g^{0\alpha}/g^{00}(1 - g^{00} g_{00})]$

$\qquad\qquad = g^{0\alpha}/g^{00}$

**[A3]** Page 4.4 $(ds/dx^0)^2 = 1/g^{00}$

In the expression for $(ds/dx^0)^2$ insert the result $dx^\alpha/dx^0 = g^{0\alpha}/g^{00}$

$(ds/dx^0)^2 = g_{00} + 2 g_{0\alpha} g^{0\alpha}/g^{00} + g_{\alpha\beta} g^{0\alpha} g^{0\beta}/(g^{00})^2$

Now $g_{0\alpha} g^{0\alpha} = g_{0k} g^{0k} - g_{00} g^{00} = \delta_0^0 - g_{00} g^{00} = 1 - g_{00} g^{00}$, and also

$g_{\alpha\beta} g^{0\alpha} g^{0\beta} = (g_{k\beta} g^{0k} - g_{0\beta} g^{00}) g^{0\beta} = (\delta_\beta^0 - g_{0\beta} g^{00}) g^{0\beta} = - g_{0\beta} g^{0\beta} g^{00} = (1 - g_{00} g^{00}) g^{00}$

Thus $(ds/dx^0)^2 = g_{00} + 2(1 - g_{00} g^{00})/g^{00} - (1 - g_{00} g^{00})/g^{00}$

$= g_{00} + (1 - g_{00} g^{00})/g^{00} = 1/g^{00}$

___

**[A4]** Page 4.7 $\beta^\alpha = \sqrt{g_{00}} g^{0\alpha}$

In the expression

$\beta^\alpha := v^\alpha/c := (dx^\alpha/dx^0)/\{\sqrt{g_{00}}[1+(g_{0\alpha}/g_{00})(dx^\alpha/dx^0)]\}$, insert

$dx^\alpha/dx^0 = g^{0\alpha}/g^{00}$ into the denominator: $\sqrt{g_{00}}[1+(g_{0\alpha}/g_{00})(dx^\alpha/dx^0)]$

$= [1+(g_{0\alpha}/g_{00})(g^{0\alpha}/g^{00})] = \sqrt{g_{00}}[1+(g_{0\alpha} g^{0\alpha})/(g_{00} g^{00})]$

$= \sqrt{g_{00}}[1+(1-g_{00} g^{00})/(g_{00} g^{00})] = \sqrt{g_{00}}[1/(g_{00} g^{00})] = 1/(\sqrt{g_{00}} g^{00})$

Then $\beta^\alpha = (g^{0\alpha}/g^{00})(\sqrt{g_{00}} g^{00}) = \sqrt{g_{00}} g^{0\alpha}$

___

**[A5]** Page 4.9 $\beta_\alpha = \lambda_{\alpha\mu} \beta^\mu = g_{0\alpha}/\sqrt{g_{00}}$

$\beta_\alpha = \lambda_{\alpha\mu} \beta^\mu = [-g_{\alpha\mu} + g_{0\alpha} g_{0\mu}/g_{00}] \sqrt{g_{00}} g^{0\mu}$

$= \sqrt{g_{00}} [-g_{\alpha\mu} g^{0\mu} + g_{0\alpha}(g_{0\mu} g^{0\mu})/g_{00}]$

$= \sqrt{g_{00}} [-(g_{\alpha k} g^{0k} - g_{\alpha 0} g^{00}) + g_{0\alpha}(1 - g_{00} g^{00})/g_{00}]$

$= \sqrt{g_{00}} [-(\delta_\alpha^0 - g_{\alpha 0} g^{00}) + g_{0\alpha}(1 - g_{00} g^{00})/g_{00}]$

$= \sqrt{g_{00}} [g_{\alpha 0} g^{00} + g_{0\alpha}/g_{00} - g_{0\alpha} g^{00}]$

$= g_{0\alpha}/\sqrt{g_{00}}$

___

**[A6]** Page 5.2 $ds^2 = 1/g^{00} (dx^0)^2 + g_{\alpha\beta} d\underline{x}^\alpha d\underline{x}^\beta$

Insert $dx^\alpha = d\underline{x}^\alpha + g^{0\alpha}/g^{00} dx^0$ into the usual expression for $ds^2$

$$ds^2 = g_{00}(dx^0)^2 + 2g_{0\alpha}dx^0[d\underline{x}^\alpha + (g^{0\alpha}/g^{00})dx^0]$$
$$+ g_{\alpha\beta}[d\underline{x}^\alpha + (g^{0\alpha}/g^{00})dx^0][d\underline{x}^\beta + (g^{0\beta}/g^{00})dx^0]$$
$$= [g_{00} + 2g_{0\alpha}g^{0\alpha}/g^{00} + g_{\alpha\beta}g^{0\alpha}g^{0\beta}/(g^{00})^2](dx^0)^2 + 2g_{0\alpha}dx^0 d\underline{x}^\alpha$$
$$+ g_{\alpha\beta}g^{0\beta}/g^{00} dx^0 d\underline{x}^\alpha + g_{\alpha\beta}g^{0\alpha}/g^{00} dx^0 d\underline{x}^\beta + g_{\alpha\beta}d\underline{x}^\alpha d\underline{x}^\beta$$

But the 2nd and 3rd last terms are identical and

$$g_{\alpha\beta}g^{0\beta}/g^{00} = [g_{\alpha k}g^{0k} - g_{\alpha 0}g^{00}]/g^{00} = -g_{\alpha 0}$$

Thus the space-time cross terms vanish. Also, $g_{0\alpha}g^{0\alpha} = 1 - g_{00}g^{00}$ and

$$g_{\alpha\beta}g^{0\alpha}g^{0\beta} = [g_{k\beta}g^{0k} - g_{0\beta}g^{00}]g^{0\beta} = -g_{0\beta}g^{0\beta}g^{00} = -[1 - g_{00}g^{00}]g^{00}$$

Finally the coefficient of $(dx^0)^2$ is

$$g_{00} + 2[1 - g_{00}g^{00}]/g^{00} - [1 - g_{00}g^{00}]/g^{00} = g_{00} + [1 - g_{00}g^{00}]/g^{00} = 1/g^{00}$$

---

[A7] Page 6.8 In the "go-with-the-flow" metric: $1/g^{00} = \rho^2 \Delta/u^4$ (see the second equation on page 5). Inverting the matrix $g_{ij}$ for the Kerr solution in which the only space-time components are the $g_{0\phi}$, one has $1/g^{00} = g_{00} - g_{0\phi}^2/g_{\phi\phi}$

Now $\rho^2 g_{00} = \rho^2 - r_s r$, $\rho^2 g_{0\phi} = r_s r a \sin^2\theta$, and $\rho^2 g_{\phi\phi} = -u^4 \sin^2\theta$. Thus

$$\rho^2/g^{00} = \rho^2 - r_s r + (r_s^2 r^2 a^2 \sin^4\theta)/u^4 \sin^2\theta$$
$$= [(\rho^2 - r_s r)u^4 + r_s^2 r^2 a^2 \sin^2\theta]/u^4$$

Inserting $u^4 = \rho^2(r^2+a^2) + r_s r a^2 \sin^2\theta$

$$1/g^{00} = = [\rho^2(r^2+a^2) + r_s r a^2 \sin^2\theta - r_s r(r^2 + a^2)]/u^4$$

But $r^2 + a^2 = r^2 + a^2 \cos^2\theta + a^2 \sin^2\theta = \rho^2 + a^2 \sin^2\theta$ and inserting this gives

$$1/g^{00} = = [\rho^2(r^2+a^2) - r_s r \rho^2]/u^4 = \rho^2(r^2 + a^2 - r_s r) = \rho^2 \Delta/u^4$$

---

[A8] Page 8.7  $4r^3 - 12 r_s r^2 - 9 r_s^2 - 8 r_s a^2 = 0$

$d/dr [(r\sqrt{\Delta} \pm r_s a)/(r - r_s)] = 0 \Rightarrow (r-r_s)[\sqrt{\Delta} + r(2r - r_s)/(2\sqrt{\Delta})] = r\sqrt{\Delta} \pm r_s a$

$$(r-r_s)[\Delta + r(2r-r_s)/2] = r\Delta \pm r_s a\sqrt{\Delta} = 0$$

$$\Rightarrow \quad r^3 - 5/2\, r^2 + 3/2\, r_s^2 r - r_s a^2 = \pm r_s a\sqrt{\Delta}$$

Inserting $\Delta = r^2 - r_s r + a^2$ and squaring both sides yields

$$4r^6 - 20r_s r^5 + 3r_s r^4 - (30r_s^2 + 8a^2)r_s r^3 + (9r_s^2 + 20a^2)r_s^2 r^2 - 12\, r_s^3 a^2 r + 4r_s^2 a^4$$
$$= 4a^2 r_s^2 (r^2 - r_s r + a^2)$$

$$4r^5 - 20r_s r^4 + 37r_s r^3 - (30r_s^2 + 8a^2)r_s r^3 + (9r_s^2 + 16a^2)r_s^2 r - 8r_s^3 a^2 = 0$$

$$(r - r_s)^2 [4r^3 - 12r_s r^2 + 9r_s^2 r - 8r_s a^2] = 0$$

Rejecting the double root at $r_s$, we have

$$4r^3 - 12r_s r^2 + 9r_s^2 r - 8r_s a^2 = 0$$